\documentclass[preprint]{aastex701}

\begin{document}

\title{Upper Limits on Pulsed Radio Emission from Unseen Compact Objects in Six Galactic Stellar Binaries}

\correspondingauthor{Fronefield Crawford}

\author{Melanie Ficarra}
\affiliation{Department of Physics and Astronomy, Franklin and Marshall College, P.O. Box 3003, Lancaster, PA 17604, USA}
\email{melanieficarra@gmail.com}

\author[0000-0002-2578-0360]{Fronefield Crawford}
\affiliation{Department of Physics and Astronomy, Franklin and Marshall College, P.O. Box 3003, Lancaster, PA 17604, USA}
\email[show]{fcrawfor@fandm.edu}

\author{T.~Joseph~W.~Lazio}
\altaffiliation{Current affiliation: Department of Climate and Space Sciences and Engineering,
  University of Michigan, Ann Arbor, MI 48109  USA} 
\affiliation{Jet Propulsion Laboratory, California Institute of Technology, 4800 Oak Grove Dr., M/S~67-201, Pasadena, \hbox{CA}  91107  USA}
\email{lazio@caltech.edu}

\begin{abstract}
We have conducted a search for radio pulsars in six Galactic stellar binary systems having unseen primary stars. All six systems have estimated primary masses in the range that could be consistent with neutron stars. We used the Green Bank Telescope at a center frequency of 350 MHz to search for dispersed periodicities and single pulses across a range of possible dispersion measures (DMs) and binary accelerations. No astrophysical signals were detected in our search. The estimated 400-MHz luminosity upper limits from the search are comparable to or smaller than the lowest values observed for almost all the known Galactic binary pulsars having cataloged 400 MHz radio luminosities. This implies that the systems we observed either do not harbor radio-emitting pulsars, contain pulsars that do not beam in our direction, or contain pulsars with luminosities that are significantly lower than this subset of the known Galactic binary pulsar population.
\end{abstract}


\section{Introduction and Background} \label{sec:intro}

Binary pulsar systems are excellent laboratories for exploring and understanding stellar evolution. Observing these systems can constrain the conditions under which a binary system remains bound after the supernova (SN) that produced the pulsar. Pulsar and companion interactions can reveal past episodes of mass transfer, including whether the pulsar underwent any recycling and, if so, how much \citep{bv91,tv23}. Further evolution information can be obtained if the mass of the companion can be measured. The survival of systems after SN events also has implications for the natal kicks imparted to pulsars during their formation. Orbital parameters provide insights into how the SN impacted the system during formation. Such systems provide the opportunity to study stellar winds of non-degenerate companions through observations of the pulsar signals \citep{ktm+96}. Alternatively, spider pulsars \citep{r13} allow for the observation of pulsar winds, which are radiatively silent unless they interact with a nearby object. The winds from the millisecond pulsar (MSP) ablate its low-mass companion, causing observable effects and constraining the pulsar’s wind and energy output due to the proximity of the companion \citep{zkr+20}. Discovery of such systems would help to refine stellar evolution models, therefore giving a more accurate prediction of the number of Galactic binary pulsars \citep{l08}, and also would help fill in important gaps in our understanding of the pulsar population as a whole. 

Following the discovery of 21 binary systems from {\it Gaia} DR3 \citep{gaia1} reported by \citet{erl+24} that contain dark stars in the neutron star (NS) mass range, there have been several related searches for radio pulsars in binaries with unseen companions \citep{glw+26,bat+26}. These studies focused on NS-main sequence binaries in which the NS may be older and radio quiet, while the search we describe here targets systems that might host MSPs which are likely to remain radio active.

Finally, if the NS accretes sufficient material and emerges as a recycled pulsar \citep{l08}, that pulsar could be incorporated into pulsar timing arrays, such as the North American Nanohertz Observatory for Gravitational Waves \citep[\hbox{NANOGrav};][]{aaa+23}. For instance, about two-thirds of the MSPs currently in the NANOGrav array are in binary systems. 

This paper reports on deep radio searches of six Galactic binary systems that may contain pulsars. Four of the systems have dark primaries orbiting subdwarf~B (sdB) companion stars. The parameters for these six systems are presented in Table~\ref{tbl-1} and are described below. The remainder of this paper discusses these six systems in further detail, and our observations, data analysis methods, and results.

\subsection{2XMM~J125556.57+565846.4}

\citet{mtf+22} identified 2XMM~J125556.57+565846.4 (hereafter 2XMM~J1255+5658) from the Large Sky Area Multi-Object Fiber Spectroscopic Telescope \citep[\hbox{LAMOST},][]{lamost2020} stellar radial velocity public database as a system with large radial velocity differences. Follow-up spectroscopy indicated it is a 2.76-day binary system consisting of a normal F-type star and a probable NS. The F-star's atmospheric and physical parameters were constrained using Bayesian analysis, which provided insights into the mass of the unseen star and the orbital inclination. Optical and UV photometry reveal ellipsoidal modulation occurring at half the orbital period due to the tidal deformation of the F-star. This constrained the mass range of the unseen primary to be between $M = 1.1$ and $2.1\,M_{\odot}$ (with a median of 1.4\,$M_{\odot}$), consistent with a NS. 2XMM~J1255+5658 was also observed to exhibit bright X-ray emission, though this X-ray emission is not coming from the unseen primary, so it is not an accretion-powered NS. The X-ray activity matches the orbit of the companion and the X-ray emission matches expectations from main sequence stars \citep[e.g.,][]{zwp20}.

\subsection{SDSS~J022932.28+713002.7}

\citet{acz+24} report the discovery of SDSS~J022932.28+713002.7 (hereafter SDSS~J0229+7130), a nascent, extremely low-mass (ELM) white dwarf (WD) orbiting a massive ($> 1\,M_{\odot}$) unseen primary with a period of 35.87\,hr. The lightcurve of the WD exhibits ellipsoidal variations, which, when combined with radial velocity data and simulations from the PHOEBE software package (used to model and analyze eclipsing binary stars; \citealt{p18}), obtains an estimate of the mass of the unseen primary of $M = 1.19^{+0.21}_{-0.14}\,M_{\odot}$. This mass suggests that it is most likely a WD near the Chandrasekhar mass or an \hbox{NS}. 

\subsection{TON~S~183} 

Subdwarf~B (sdB) stars are interpreted as having a thin hydrogen layer surrounding a helium-burning core \citep{hpmmi02,hpmm03,h16}.  From an evolutionary perspective, they are considered to be post-main sequence stars on the extreme horizontal branch, and all formation channels involve binary evolution \citep{ycl+18,ycc+20}.  If the companion to an sdB star is an \hbox{NS}, it potentially would have accreted sufficient mass to emerge as a recycled pulsar, which was a motivation for searching for radio pulsations from this and the other three sdB stars in our sample. This sdB star was identified by \cite{ghp+10} as having a primary with a mass of~$0.94^{+0.39}_{-0.31}\,M_\odot$. While they classify the primary as most likely being a WD, the range of estimated masses does extend above the Chandrasekhar limit. \cite{mce+11} find no X-ray emission from this system, using the Neil Gehrels Swift Observatory, from which they derive an upper limit on the wind of the sdB star.

\subsection{HE~0929$-$0424}\label{sec:he0929-0424}

This sdB star was identified by \citet{ghp+10} as having an unseen primary with a mass of~$1.82^{+0.88}_{-0.64}\,M_\odot$. The range of allowed masses is sufficiently large that they provide no classification for the nature of the companion. \cite{mce+11} find no X-ray emission from this system, using the Neil Gehrels Swift Observatory, from which they derive an upper limit on the wind of the sdB star.

\subsection{PG~1232$-$136}\label{sec:pg1232-136}

This sdB star was identified by \citet{ghp+10} as having a primary with a mass of at least 6\,$M_\odot$.  While this limit is well above the expected maximum mass of a NS, it also depends upon the assumed mass of the sdB star itself.  Certain mass transfer scenarios for this system could result in the sdB star having a mass larger than the canonical value of~0.5\,$M_\odot$ \citep{h86}, which would lower the estimated mass of the companion.  An independent mass for the sdB star, which might be ultimately obtained from {\it Gaia} astrometry \citep[e.g.,][]{cgrpgfv22}, could provide additional constraints on the mass of the companion.  However, we are not aware of any such independent mass estimate, and none was available at the time of the observations reported here. \cite{mce+11} find no X-ray emission from this system, using the Neil Gehrels Swift Observatory, from which they derive an upper limit on the wind of the sdB star. 

\subsection{BPS~BS~16981$-$0016}\label{sec:bpsbs1691-0016}

This sdB star was identified by \citet{gdd+23} as having a primary with a mass of~$1.50^{+0.37}_{-0.45}\,M_{\odot}$, which they suggest as being either a massive WD or a NS. If the primary is a WD, this system could represent the double degenerate model, which is a theoretical progenitor class of Type~Ia SNe that is estimated to form approximately 9\% of sdBs \citep{snt+15}.

\section{Observations}\label{sec:observe}

We have used the Green Bank Telescope (GBT) in Green Bank, West Virginia to search these six binary systems for pulsed radio emission, which would signal that the unseen primaries are radio pulsars. In all observations, we used a center observing frequency of 350\,MHz, with a 100\,MHz bandwidth split into 4096 channels of width 24.4\,kHz that were sampled every 81.92\,$\mu$s. These are the same observing parameters as used in the Green Bank Northern Celestial Cap (GBNCC) survey \citep{slr+14}, but we used much longer integration times. Since pulsars typically have steep spectra and the expected dispersion smearing and scattering is relatively small for these sources (see below), this low observing frequency is appropriate for this search. 

We observed the six targets in two widely separated observing sessions. The first session occurred on 2011 May~6 and~2011 May~9 and used the Green Bank Ultimate Pulsar Processing Instrument (GUPPI) backend \citep{drd+08}. We observed the following (sdB) stars (with durations): TON~S~183 (60\,min), HE~0929$-$0424 (60\,min), and PG~1232$-$136 (55\,min). The second session occurred on 2024 January~16 and 2024 February~11, using the Versatile GBT Astronomical Spectrometer (VEGAS) backend \citep{pbb+15}, which replaced GUPPI. We observed the following stars (with durations): 2XMM~J1255+5658 (45\,min), BPS~BS~16981$-$0016 (75\,min), and SDSS~J0229+7130 (75\,min). Table~\ref{tbl-2} summarizes the properties of these searches.

\subsection{Prior Observations and Analysis of Some of the Targets}

We note that our observations from 2011 (i.e., those targeting HE 0929$-$0424, PG 1232$-$136, and TONS S 183) repeat similar, earlier observations taken of these same three sdB targets with the GBT. These prior observations were searched and published by \citet{cvs11}. Our observations and analysis are different in some ways, and so we explore parameter space that was not searched by these authors. Specifically, they used half the bandwidth we did and had shorter integrations (15 min in the cases of HE 0929-0424 and PG 1232-136), resulting in reduced raw sensitivity. Their acceleration range searched was not specified, and they did not do a separate segmented chunk search, so they are not likely to be as sensitive to the highest expected accelerations as we are. In addition, they kept the top 30 candidates in each search, but no S/N cutoff was specified. It is possible that this candidate list did not extend down to our S/N cutoff of 7. For their single pulse search, they did not specify a maximum width to which they were sensitive (i.e., a maximum box car filter width used), and no minimum S/N was specified. Thus, our search is likely to be more sensitive to pulsars occupying certain areas of parameter space.

In addition, \citet{ovm+20} processed two of the observations we took in 2011 (HE 0929$-$0424 and PG 1232$-$136). Many of their processing parameters were similar to ours (e.g., they used the PRESTO processing package, had a similar DM search range, and conducted a segmented chunk search with acceleration as well as a single-pulse search). However, their processing differed from ours in several important ways. They assumed a primary NS mass of $1.4 M_{\odot}$ in each case, whereas we used more specific estimates for the masses of the primaries. This led to different maximum accelerations searched, and ours extended to higher values. They used a minimum S/N of 8 for candidates retained from the PRESTO processing, while we kept and folded all candidates produced, and inspected those refolded candidates with  S/N~$> 7$. For their single pulse search, they used PRESTO’s single pulse script, with a minimum S/N of 8 and a maximum boxcar filtering width that maintained full sensitivity to pulse widths up to 20 ms. In our processing, we used the HEIMDALL \citep{bbb+12} and FETCH \citep{aab+20} single pulse search and classification packages as well as PRESTO. For our HEIMDALL/FETCH search we maintained full sensitivity to widths up to 42 ms, twice the \citet{ovm+20} value, and we used a minimum S/N of 7 for both types of searches.

Of our six targets, three (2XMM J1255+5658, BPS BS 16981$-$0016, and SDSS J0229+7130) were not analyzed by either \citet{cvs11} or \citet{ovm+20} and so they represent completely new searches. 

\section{Data Analysis and Results}

\subsection{Periodicity Search}

We first searched the data for periodicities using the PRESTO software package \citep{r01, rem02}. We cleaned the data of RFI using the rfifind PRESTO tool. For the observations taken in~2011, $\sim 1$\% of the data were masked in this process, while for the later observations in~2024, between~16\% and~18\% of the data were masked in each case. For these latter observations, the majority of the flagged RFI was persistent over the course of the observation and occurred in four distinct subbands between~360 and~380\,MHz. After RFI masking, we dedispersed the raw data at a total of 11565 DM trials between~0 and~214\,pc\,cm$^{-3}$. The DM spacing was determined by the PRESTO DDplan.py script. This DM range encompassed the maximum Galactic DM ranges for all six sources according to both the YMW16 \citep{ymw17} and NE2001 \citep{cl02} models. Table~\ref{tbl-2} provides full search details for all six systems.

For each dedispersed time series, we searched a range of possible accelerations corresponding to a maximum Fourier bin drift of $z_{\rm max} = 1200$ over the course of the observation. The corresponding maximum acceleration to which full sensitivity was maintained in the search was
\begin{equation}
a_{\rm max} = \frac{z_{\rm max} c P}{h T^{2}},
\label{eqn-1}
\end{equation}
where $c$ is the speed of light, $P$ is the fundamental spin period of the pulsar, $h$ is the number of harmonics used, and $T$ is the integration time \citep[e.g.,][]{br18}. For a putative 1\,ms spin period and eight harmonics, the maximum acceleration searched was a few m\,s$^{-2}$ in each case (Table~\ref{tbl-2}). This acceleration limit increases linearly with spin period, so longer-period pulsars would have larger acceleration ranges encompassed by the search. We note that these search limits for a 1\,ms pulsar exceed the maximum expected binary accelerations for three of our systems. These accelerations are calculated below and listed in Table~\ref{tbl-1}.

The maximum acceleration for a pulsar in a circular orbit ($e=0$) seen edge-on is presented by \citet[Eq.~1]{sbc+21}. Using the treatment of \citet{fkl01}, we can include the eccentricity $e$ and modify the expression in \citet{sbc+21} to be
\begin{equation}
a_{\rm max} = \left( \frac{GM_{c}^{3}}{M_{\rm tot}^{2}}    \right)^{1/3} \left( \frac{2\pi}{P_{b}} \right)^{4/3} \frac{(1+e)^{2}}{(1-e^{2})}, 
\end{equation}
where $M_{c}$ is the mass of the companion, $M_{\rm tot}$ is the total mass of the pulsar plus the companion, and $P_{b}$ is the binary orbital period.
Table~\ref{tbl-1} contains the specific details for the six systems and the calculated maximum expected accelerations. For three of our targets, our acceleration search range encompassed these maximum expected accelerations for all reasonable spin periods, indicating that Doppler smearing is not a factor reducing sensitivity. For the other three systems, the expected maximum accelerations in Table~\ref{tbl-1} significantly exceed our search range, but the segmented chunk search (see below) extended this to a much larger range that encompassed these maximum values.

Note that in each case, the integration time was a small fraction (less than about~10\%) of the system's orbital period (Table~\ref{tbl-1}), which indicates an approximately linear drift in the pulsar's spin frequency owing to binary acceleration during the course of the observation \citep[e.g.,][]{rce03}. As a test of the processing pipeline, the bright (isolated) pulsar PSR~B1508+55 was clearly detected in a 60\,s diagnostic observation. However, no promising periodic signals were detected in the observations of the six targets.

Candidates that were identified in the spectra were refolded at the candidate periods, period derivatives, and DMs, while being allowed to vary slightly in this process within a small range of parameters to maximize the S/N in the refold. All of the resulting folded candidates having S/N greater than 7 were inspected visually, but none showed promise.

We also conducted a search of smaller segments of the full integration in order to increase our search acceleration range, which could be important for the three targets which might exceed our acceleration search range (Table~\ref{tbl-1}). For each of the six targets, we divided the integrations into 660\,s (11\,min) chunks (which were staggered and overlapped each other by~50\%) and processed them using the same set of dedispersions and the same maximum frequency bin drift of 1200 as used above. This reduced the raw sensitivity by a factor of 2--2.5, depending on the integration time of the original observation, but it greatly expanded the  acceleration search range. Substituting the 660\,s chunk integration time into Eq.~\ref{eqn-1} gives a new $a_{\rm max} = 103.3$ m s$^{-2}$ (for a 1\,ms pulsar) in each case. This is comparable to the largest maximum expected acceleration of our targets (Table \ref{tbl-1}). Searching segments of the integrations separately also enhances sensitivity to scintillating, eclipsing, or nulling pulsars that may be visible for only part of the full integration. This may be important since our target systems are binaries, and pulsars could experience orbital effects that enhance or reduce emission during certain phases of the orbit. As above, all candidates identified in the spectra were refolded and inspected visually, but none were classified as likely real.

In addition to this, we conducted a search for longer periodicities using the fast-folding algorithm (FFA) algorithm in the RIPTIDE software package \citep{mbs+20}. Our FFA search of the dedispersed data files was sensitive to periods between 1 and 30 s. Pulsars in binary systems with non-degenerate companions are likely to be at least partially recycled if mass transfer occurred, and so we would not expect them to have the very long spin periods that result from extended spin-down \citep{bv91}. In addition, acceleration from binary motion was not accounted for in this FFA search. No significant FFA signals were detected above S/N of 7, except for strong detections of the test pulsar PSR B1508+55, which appeared as sub-harmonics of its fundamental 0.740-s spin period.

\subsection{Single Pulse Search}

We searched the data for dispersed single pulses using the HEIMDALL and FETCH analysis packages. HEIMDALL \citep{bbb+12,bj24} identified impulsive signals in the dedispersed time series for trial DMs ranging from 0 to 500 pc cm$^{-3}$. The DM spacing was determined by HEIMDALL. Each dedispersed time series was searched using boxcar matched filter widths of between 1 and 512 samples, in successive powers of 2. This ensured sensitivity to pulse widths up to 42 ms. All candidates with a S/N above 7 were retained for analysis by FETCH. FETCH is a neural network classifier that distinguishes between astrophysical signals and RFI using morphological characteristics \citep{aab+20}. FETCH assigned a probability of being real to each candidate from HEIMDALL that was passed to it. A total of 40 strong candidates from the 60-s diagnostic observation of the bright pulsar PSR B1508+55 were detected and identified with confidence as real, indicating that the processing pipeline was working properly. Only one candidate from the six target observations was rated by FETCH above 50\% likelihood of being real (with a rating of 51\%), but upon visual inspection it was clearly strong RFI that had been misidentified.

In addition, we independently searched for single pulses with S/N above 7 using PRESTO. We searched DMs from 0 to 214 pc cm$^{-3}$ (the same range as the periodicity search) with the default set of boxcar matched filter widths from PRESTO of 1, 2, 3, 4, 6, 9, 14, 20, and 30 samples. For the diagnostic observation of PSR B1508+55, many strong pulses were detected (with S/N between 10 and 40). In the searches of the six targets, several marginal candidates were detected at large DMs (greater than 100 pc cm$^{-3}$), but these were not confirmed in the HEIMDALL and FETCH search output. The large DMs of these candidates also suggest they are not likely to be associated with the targeted sources, if they are real. 

\section{Discussion}

\subsection{Expected Dispersion Smearing, Scattering, and Scintillation Effects}

\begin{figure}
\begin{center}
\includegraphics[width=1.0\textwidth]{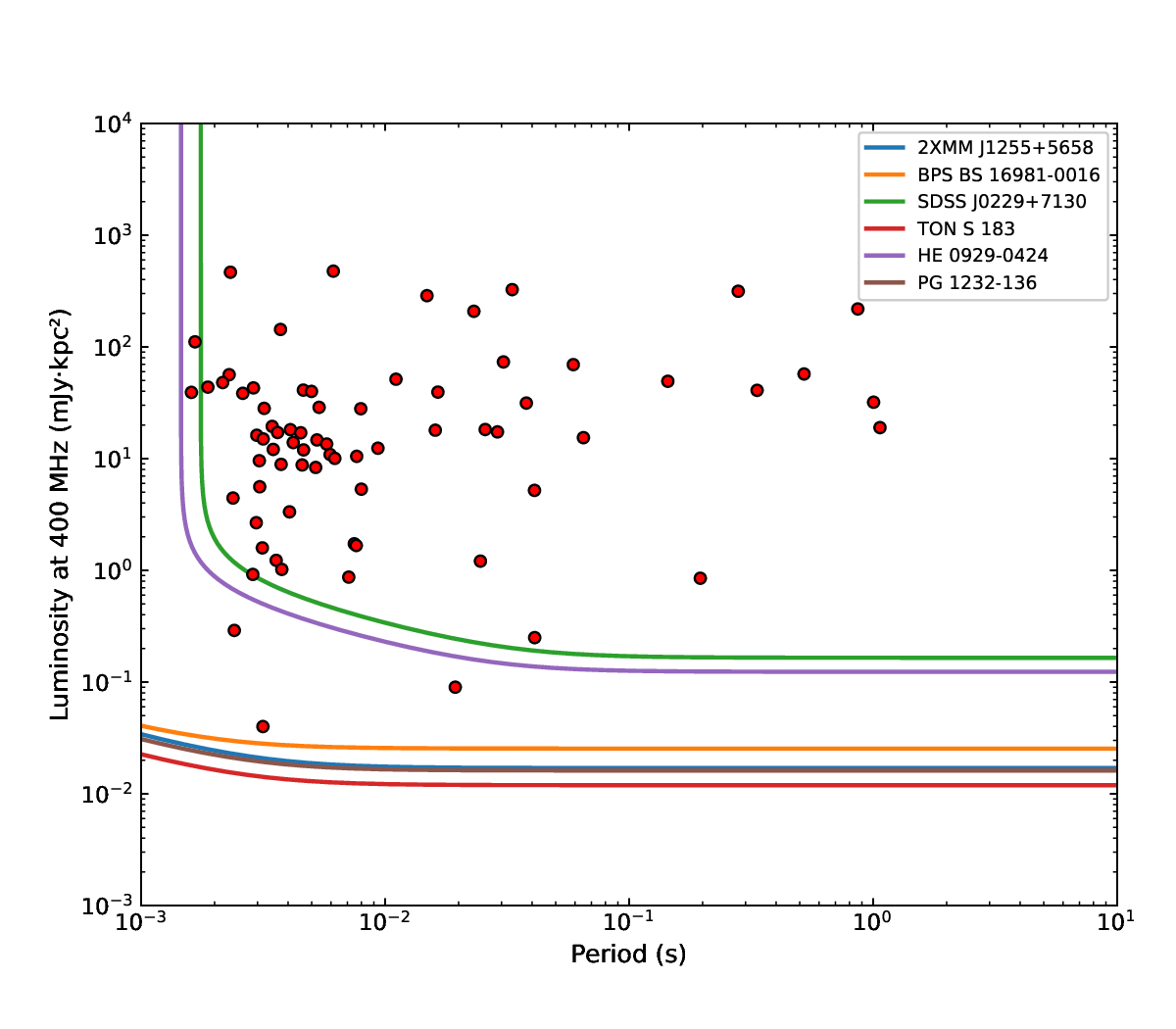}
\end{center}
\caption{Estimated 400-MHz luminosity limits vs. spin period for our six targets, plotted as differently colored curves. The population of known Galactic binary radio pulsars with measured 400-MHz luminosities from the ATNF catalog \citep{mht+05} is plotted as red dots. For the calculation of the sensitivity curves, we used the estimated DM for each source, derived from its distance and the NE2001 Galactic electron model \citep{cl02} (see Table \ref{tbl-2}). We also used the NE2001 model to estimate the scattering times for these targets at their estimated distances (see Table \ref{tbl-2}). The calculated sensitivity limits assumed an intrinsic width corresponding to a pulsed duty cycle of 5\%, with the sampling time, DM smearing time within channels, and scattering time added to this width in quadrature to obtain an effective pulse width.  For all but two of the targets, the searches were sensitive to the lowest luminosities observed for this set of Galactic binary radio pulsars. The searches of the other two targets would be sensitive to most of these systems.}
\label{fig-1}
\end{figure}

Using the NE2001 Galactic electron model \citep{cl02} and the estimated distances of the six systems, we estimated the DMs of the six targets and the resulting dispersive smearing within channels. These are listed in Table \ref{tbl-2}. For all six targets, these smearing times are less than 0.2 ms. For four targets, the expected scattering at 350 MHz is also less than 0.2 ms (Table \ref{tbl-2}), which preserves sensitivity to MSPs. For SDSS~J0229+7130 and HE 0929$-$0424, the expected scattering times of 1.74 and 1.44 ms significantly degrade sensitivity to MSPs, but otherwise do not affect sensitivity to longer-period pulsars. The total smearing can be approximated as the quadrature sum of the intrinsic pulse width (which we have assumed to be 5\% of the pulse period), the sampling time, the scattering time, and the DM smearing time within channels. This effective pulse width was then used to calculate the search sensitivity as a function of pulse period, as shown in Fig. \ref{fig-1}. As seen in this figure, we retain good sensitivity to periods greater than a few milliseconds for all six targets. Using the NE2001 model, we also calculated the expected scintillation times and bandwidths at 350 MHz for all six sources (see Table \ref{tbl-2}). In all cases these were much less (i.e., at least an order of magnitude smaller) than the integration times and the observing bandwidth used, indicating that scintillation did not significantly affect pulsar detectability in any of these observations. 

\subsection{Sensitivity Limits}

For three of the systems (2XMM~J1255+5658, BPS~BS~16981$-$0016, and SDSS~J0229+7130), the best prior limits on the flux densities of pulsars were from the GBNCC survey and were 1.2\,mJy at 350\,MHz for pulse periods greater than about 5 ms \citep{slr+14}. We searched these targets more deeply by increasing the integration time while using the same telescope and observing setup. Our upper limits on the flux density at 350 MHz improve upon those from the GBNCC survey by a factor of approximately 20. 

As described above, the other three systems (TON~S~183, HE~0929$-$0424, and PG~1232$-$136) were previously observed by \citet{cvs11}. Our sensitivity limits are several times better than those listed in their Table 3. Also, \citet{ovm+20} processed our data taken on two of these three targets (HE~0929$-$0424 and PG~1232$-$136) and found nothing.  Our sensitivity limits are comparable to the limits cited by \citet{ovm+20} in their Table 1 when differences in assumptions in the sensitivity calculations are accounted for (e.g., the receiver temperature used).

Our integration times in each case were long enough that the luminosity limits reach down to the lowest values of the cataloged 400 MHz luminosities of the Galactic binary pulsar population (Fig. \ref{fig-1}). To get these 400-MHz luminosity limits, we first calculated the minimum flux density at 350 MHz using the modified radiometer equation \citep{dtw+85}. We first computed the system temperatures for our six targets at 350 MHz, derived from the sky maps of \citet{hss+82} and the receiver temperature of 23 K quoted in the GBNCC survey paper \citep{slr+14}. We used a minimum S/N of 7, a telescope gain of 2 K/Jy for the GBT, two polarizations, a 100 MHz bandwidth, and the specific integration time for each target.\footnote{See, e.g., \url{https://gbtdocs.readthedocs.io}.} We assumed a 5\% intrinsic pulsed duty cycle and no digitization loss in the sampling. The dispersive smearing within channels and the scattering times (listed in Table \ref{tbl-2}) were included in the pulse width by adding these in quadrature to the intrinsic width and sampling time. We did not  consider the fraction of data masked as RFI in the calculation. These parameters and calculated flux density limits are presented in Table \ref{tbl-2}. 

The 350-MHz flux density limits were then scaled to 400 MHz using an assumed spectral index of $\alpha = -1.6$, an average for the pulsar population (e.g., \citealt{jvk+18}). The 400 MHz (pseudo)luminosity limits were then found by multiplying each 400 MHz flux density limit by the square of the estimated distance to each system. These 400 MHz luminosity limits are presented in Table~\ref{tbl-2}.

\subsection{Implications for Long-Period Transients and White Dwarf Radio Emitters}

While the initial focus of these observations was to search for radio pulsars in these systems, the observations were also sensitive to periodic radio emission from other sources. In particular, long-period transients \citep[LPTs,][]{rhc+26} are an emerging class of radio transients with some characteristics similar to those of pulsars, such as periodic, highly-polarized emission. However, notably distinct from the pulse periods for pulsars (of order 1.5\,{ms} to about~8\,s), long-period transients have pulse periods of seconds to hours.

Of particular relevance to these observations is that at least some LPTs appear to be the result of pulsed radio emission from WDs or systems containing WDs. As discussed in~\S\ref{sec:intro}, for some of the systems, WDs remain candidates for the unseen objects, and thus our limits on pulsed radio emission can be interpreted in terms of radio emission from WDs, or LPTs more generally.

The durations of our observations (approximately 1\,hr) do limit what constraints can be placed on LPTs in any of these systems.  If we require several full cycles in order to identify periodic emission, we would detect only those LPTs with pulse periods shorter than approximately 15\,min. Approximately 75\% of the current sample of LPTs have pulse periods longer than 15\,min, though the two so-called WD pulsars, \objectname{AR~Sco} \citep{mgh+16} and \objectname{eRASSU~J1912$-$4410} \citep{pmb+23}, both have pulse periods shorter than about 5\,min. Furthermore, many LPTs show intermittent activity, with the intervals for some objects being shorter than 1\,month. Thus, the absence of any radio emission from our observations may not provide strong constraints on any pulsed radio emission from WDs or LPTs.

Our flux density limits place severe constraints on the existence of any WD pulsars in any of these systems. The faintest LPT in the current sample is \objectname{CHIME~J0630+25} \citep{dcc+25}, with a range of characteristic flux densities of~0.4--1.9\,mJy, which is at least a factor of five above our detection limits at a similarly low observing frequency. Many of the LPTs have chacteristic flux densities of order 100\,mJy or larger. Crucially, both of the WD pulsars have flux densities of order 10\,mJy, meaning that analogous WDs would have been detected in our analysis. Within the limitations of our observations, we conclude that, if any of these systems contain WDs instead of NSs, the WDs do not pulse at radio wavelengths.

\subsection{Comparison to the Known Galactic Radio Binary Population}

Figure~\ref{fig-1} compares our luminosity limits to the known population of 72 Galactic radio binary pulsar systems listed in the ATNF Pulsar Catalog\footnote{\url{https://www.atnf.csiro.au/research/pulsar/psrcat/}} \citep{mht+05} with measured 400-MHz luminosities. The sensitivity curves for four of the targets indicate that all of these Galactic binary pulsars would be sufficiently luminous to be detectable in our search if they were placed at our target distances. For our other two targets, all but three of the Galactic systems would have likely been detectable in our observations. One of these three low-luminosity pulsars is \objectname{PSR J1745$-$0952}. Its derived distance from its DM of 64.3\,pc cm$^{-3}$ shows a significant discrepancy between the values of 0.23\,kpc and 1.84\,kpc from the YMW16 \citep{ymw17} and NE2001 \citep{cl02} models, respectively. The YMW16 value is adopted by the ATNF Pulsar Catalog and gives a subsequent luminosity that smaller by a factor of 66 than the value derived from the NE2001 model; this may give an artificially small luminosity, and the system may in fact lie above our luminosity detectability threshold. Similarly, the other two pulsars, \objectname{PSR~J0307+7443} and \objectname{PSR~J1124$-$3653}, show discrepancies in their estimated distances using the two Galactic electron models. In these cases, the discrepancies are not as significant: adopting the NE2001 model would increase the implied luminosity of the pulsar by a factor of~2 and 3, respectively.   

For \objectname{2XMM~J1255+5658}, the non-detection is consistent with the NS being dormant. However, for an assumed 20\% beaming fraction \cite[e.g.,][]{l08}, there is an 80\% chance that there is an active pulsar in the system which is beaming away from us. For SDSS~J0229+7130, we cannot distinguish between the WD and~NS hypotheses for the dark primary star. For the four primary stars orbiting sdB companions, if we assume a 20\% beaming fraction for each, then there is about a 40\% chance that all of the pulsars are beaming away from us and we would not detect any pulsar components, even if all of the dark objects were active pulsars. Thus, the constraints on the emission must include the caveat that there is a reasonable chance that beaming is the reason for our non-detections.

\section{Conclusions}

We conducted extended GBT search observations at 350 MHz for pulsed radio emission from six Galactic binary systems that might contain previously undetected radio pulsars. We detected no periodic or dispersed impulsive radio signals from any of the six targets. Our flux density limits improve upon the GBNCC survey limits by a factor of about 20. Our luminosity limits from these observations are comparable to or lower than the lowest luminosities of almost all of the known Galactic binary radio pulsars having measured 400 MHz luminosities. If any of the targeted systems do contain pulsars, they are either not beaming in our direction (a reasonable possibility assuming a typical beaming fraction), or they are significantly less luminous than the known Galactic binary pulsar population. 

\begin{acknowledgments}
We thank Jay Frothingham for valuable support and assistance with the GBT observations. This research has made use of NASA’s Astrophysics Data System and the Australian Telescope National Facility (ATNF) Pulsar Catalogue. The Green Bank Observatory is a facility of the National Science Foundation operated under cooperative agreement by Associated Universities, Inc. Part of this research was carried out at the Jet Propulsion Laboratory, California Institute of Technology, under a contract with the National Aeronautics and Space Administration. This work was supported in part by National Science Foundation (NSF) Physics Frontiers Center award Nos.~1430284 and 2020265, and used the Franklin and Marshall College compute cluster, which was funded through NSF grant 1925192.
\end{acknowledgments}

\vspace{5mm}
\facilities{Green Bank Telescope}

\software{HEIMDALL \citep{b12,bbb+12}, FETCH 
\citep{aab+20}, PRESTO \citep{r01,rem02}
          }


\begin{deluxetable}{lccc}
\tablecaption{Parameters of Six Observed Galactic Binary Systems}
\tablewidth{0pt}
\tabletypesize{\footnotesize}
\setlength{\tabcolsep}{4pt}
\renewcommand{\arraystretch}{0.95}
\tablehead{
\colhead{Target} \vspace{-0.6cm} &  \colhead{2XMM J1255+5658} & \colhead{BPS BS 16981$-$0016} & \colhead{SDSS J0229+7130} \\
\colhead{} & \colhead{} & \colhead{} & \colhead{} 
}
\startdata
Right ascension (J2000)                              & 12:55:57  &    14:33:31 &  02:29:32 \\
Declination (J2000)                                  & +56:58:45 & $-$01:14:43 & +71:30:00 \\
Observation MJD                                      & 60325     & 60351       & 60351     \\
Companion star possible type                         & F-star    & subdwarf B-star & ELM WD \\
Mass of companion star, $M_{c}$ ($M_{\odot}$)        & 1.13      &  0.5        & 0.018    \\
Expected mass of unseen primary star ($M_{\odot}$)   & $1.4_{-0.3}^{+0.7}$  &  $1.50_{-0.45}^{+0.37}$  & $1.19_{-0.14}^{+0.21}$    \\ 
Orbital period, $P_{b}$ (d)                          & 2.76      & 892.5       & 1.5 \\
Orbital eccentricity, $e$                            & $\approx 0$             & 0.36       &  ?   \\ 
Maximum expected acceleration (m s$^{-2}$)           & 3.0       &  0.004      & 0.127$^{a}$ \\
Reference                                            & [1]       & [2]         & [3] \\
\hline
Target & TON~S~183 & HE~0929$-$0424 & PG~1232$-$136 \\
\hline
Right ascension (J2000)                              & 01:01:17    & 09:32:02    &  12:35:18   \\
Declination (J2000)                                  & $-33$:42:45 & $-$04:37:37 & $-13$:55:09 \\
Observation MJD                                      & 55690       & 55687       & 55687       \\
Companion star possible type                         & subdwarf B-star & subdwarf B-star & subdwarf B-star \\
Mass of companion star, $M_{c}$ ($M_{\odot}$)         & 0.5         &  0.5        & 0.5 \\
Expected mass of unseen primary star ($M_{\odot}$)   & $0.94^{+0.39}_{-0.31}$  & $1.82^{+0.88}_{-0.64}$  & $> 6$? \\
Orbital period, $P_{b}$ (d)                           & 0.83        & 0.44        & 0.36 \\
Orbital eccentricity, $e$                             & 0.052       & $\approx 0$ & 0.06 \\
Maximum expected acceleration (m s$^{-2}$)           & 16.41       & 48.23       & 117.01$^{b}$ \\
Reference                                            & [4]         & [4]         & [4] \\
\enddata 
\tablecomments{The mass for sdB stars is assumed to be 0.5\,$M_\odot$ \citep{h86}. $^{a}$The maximum acceleration reported here assumes an orbital eccentricity of 0. 
$^{b}$Assumes a mass of the primary star of 6\,$M_\odot$. References: [1] \citet{mtf+22}; [2] \citet{gdd+23}; [3] \citet{acz+24}; [4] \citet{ghp+10}.} 
\label{tbl-1}
\end{deluxetable}

\begin{deluxetable}{lccc}
\tablecaption{Six Galactic Binary Systems Searched}
\tablewidth{0pt}
\tabletypesize{\footnotesize}
\setlength{\tabcolsep}{4pt}
\renewcommand{\arraystretch}{0.95}
\tablehead{
\colhead{Target} \vspace{-0.6cm} &  \colhead{2XMM J1255+5658} & \colhead{BPS BS 16981$-$0016} & \colhead{SDSS J0229+7130} \\
\colhead{} & \colhead{} & \colhead{} & \colhead{} 
}
\startdata
Estimated distance (kpc)                  & 0.60      & 0.73      & 1.66  \\
Estimated DM (pc cm$^{-3}$)               & 12        & 9         & 40    \\ 
Maximum DM along line of sight (pc cm$^{-3}$)                 & 23/31     & 27/30     & 190/130 \\
Dispersive channel smearing for expected DM (ms)              & 0.057      & 0.041      & 0.19 \\ 
Scattering time at 350 MHz (ms)           & 0.135     & 0.056     & 1.74  \\
Scintillation bandwidth at 350 MHz (MHz)  & 0.138     & 0.333     & 0.011 \\
Scintillation time at 350 MHz (s)         & 209       & 360       & 97    \\
Search integration time (min)             & 45        & 60        & 75 \\
Estimated sky temperature at 350 MHz      & 32        & 41        & 67 \\
Estimated system temperature (K)          & 55        & 64        & 90 \\
Maximum acceleration searched, $a_{\rm max}$ (m s$^{-2}$) & 6.2 & 3.5 & 2.2 \\
350 MHz flux density limit (mJy)          & 0.060     & 0.060     & 0.076 \\
400 MHz luminosity limit (mJy kpc$^{2}$)  & 0.017     & 0.025     & 0.166 \\ 
\hline
Target & TON~S~183 & HE~0929$-$0424 & PG~1232$-$136 \\
\hline
Estimated distance (kpc)                  & 0.54      & 1.90  & 0.57  \\
Estimated DM (pc cm$^{-3}$)               & 12        & 39    & 12    \\ 
Maximum DM along line of sight (pc cm$^{-3}$) & 19/29 & 40/51 & 29/39 \\
Dispersive channel smearing for expected DM (ms) & 0.06 & 0.18  &  0.06 \\ 
Scattering time at 350 MHz (ms)           & 0.113 & 1.44  & 0.117     \\
Scintillation bandwidth at 350 MHz (MHz)  & 0.164 & 0.013 & 0.159     \\
Scintillation time at 350 MHz (s)         & 216   & 114   & 219       \\
Search integration time (min)             & 60    & 60    & 55        \\
Estimated sky temperature at 350 MHz      & 32    & 23    & 41        \\
Estimated system temperature (K)          & 55    & 46    & 64        \\
Maximum acceleration searched, $a_{\rm max}$ (m s$^{-2}$)  &  3.5  &  3.5 & 4.1  \\
350 MHz flux density limit (mJy)          & 0.052 & 0.043 & 0.063     \\
400 MHz luminosity limit (mJy kpc$^{2}$)  & 0.012 & 0.123 & 0.016     \\
\enddata 
\tablecomments{The maximum Galactic DM along the line of sight was estimated from the YMW16 \citep{ymw17} and NE2001 \citep{cl02} models, respectively. The target DM and scintillation parameters were estimated from the reported distance using the NE2001 model. The estimated sky temperature is from \citet{hss+82} scaled to 350 MHz, and the system temperature is the receiver temperature (23 K; \citealt{slr+14}) plus the sky temperature. The maximum acceleration searched assumed a spin period of 1 ms and increases linearly with spin period. The 400-MHz luminosity limit was determined by scaling the 350 MHz flux density limit to 400 MHz and computing the pseudoluminosity by multiplying by the distance squared. These luminosity limits apply for long pulsar periods (see Fig. \ref{fig-1}).} 
\label{tbl-2}
\end{deluxetable}

\end{document}